\documentclass[sigconf]{acmart}
\renewcommand\footnotetextcopyrightpermission[1]{} 
\usepackage[utf8]{inputenc}
\usepackage{natbib}
\usepackage{graphicx}
\usepackage[caption=false]{subfig}
\usepackage{multirow}
\usepackage{multicol}
\usepackage{array}
\usepackage{xcolor}
\usepackage{float}
\usepackage[normalem]{ulem}
\usepackage{listings}
\usepackage{makecell}
\usepackage{hyperref}
\usepackage{xspace}

\newcommand{\dcdb}{DCDB\xspace}

\title[Monitoring with \dcdb]{From Facility to Application Sensor Data:\\ Modular, Continuous and Holistic Monitoring with \dcdb}

\author{Alessio Netti}
\email{alessio.netti@lrz.de}
\affiliation{Leibniz Supercomputing Centre}
\affiliation{Technical University of Munich}
\author{Micha M\"uller}
\email{micha.mueller@lrz.de}
\affiliation{Leibniz Supercomputing Centre}
\affiliation{Technical University of Munich}
\author{Axel Auweter}
\email{axel.auweter@megware.com}
\affiliation{Megware GmbH}
\author{Carla Guillen}
\email{carla.guillen@lrz.de}
\affiliation{Leibniz Supercomputing Centre}
\author{Michael Ott}
\email{michael.ott@lrz.de}
\affiliation{Leibniz Supercomputing Centre}
\author{Daniele Tafani}
\email{daniele.tafani@lrz.de}
\affiliation{Leibniz Supercomputing Centre}
\author{Martin Schulz}
\email{schulzm@in.tum.de}
\affiliation{Technical University of Munich}

\copyrightyear{2019}
\acmYear{2019}
\acmConference[SC '19]{The International Conference for High Performance Computing,
Networking, Storage, and Analysis}{November 17--22, 2019}{Denver, CO, USA}
\acmBooktitle{The International Conference for High Performance Computing, Networking,
Storage, and Analysis (SC '19), November 17--22, 2019, Denver, CO, USA}
\acmPrice{15.00}
\acmDOI{10.1145/3295500.3356191}
\acmISBN{978-1-4503-6229-0/19/11}

\begin{document}

\begin{abstract}
Today's HPC installations are highly-complex systems, and their complexity will only increase as we move to exascale and beyond. At each layer, from facilities to systems, from runtimes to applications, a wide range of tuning decisions must be made in order to achieve efficient operation. This, however, requires systematic and continuous monitoring of system and user data. While many insular solutions exist, a system for holistic and facility-wide monitoring is still lacking in the current HPC ecosystem.

In this paper we introduce \dcdb, a comprehensive monitoring system capable of integrating data from all system levels. It is designed as a modular and highly-scalable framework based on a plugin infrastructure. All monitored data is aggregated at a distributed noSQL data store for analysis and cross-system correlation. We demonstrate the performance and scalability of \dcdb, and describe two use cases in the area of energy management and characterization.
\end{abstract}


\maketitle
\renewcommand{\shortauthors}{Netti et al.}

\section{Introduction}
\label{sections:introduction}

As the use of High-Performance Computing (HPC) technologies in science and industry continues to increase and the race for the world’s fastest machines yields ever-growing sizes of supercomputers, the complexity across all layers, from the data center building infrastructure, over the HPC system's hard- and software, to the scientific application software, has increased significantly. Current HPC systems consume power in the same order as large industrial facilities, and future \emph{exascale} systems are expected to pose an even greater challenge both in terms of energy efficiency~\cite{villa2014scaling} and resilience~\cite{cappello2014toward}, in addition to exposing more and more parallelism in highly heterogeneous architectures~\cite{osti_1473756}. To cope with these complexities and in order to retain insight into the working conditions of a system, administrators and users alike rely on monitoring tools that collect, store and evaluate relevant operational, system and application data.

With the adoption of new technologies, such as liquid cooling, the appearance of heterogeneous and accelerated systems, the increased use of dynamic tuning mechanisms at all system levels, and the establishment of complex workflows, one can observe the necessity for tighter integration of HPC systems and applications with their surrounding data center as well as with both local and global resource management. As a consequence, it is imperative that we (a) gather  application, system and facility data, (b) provide the mechanisms to efficiently manage and store them, and (c) establish the foundation to integrate the data across all layers. As of today, though, distinct monitoring systems for the data center infrastructure, system hardware, and application performance are commonly used~\cite{gimenez2017scrubjay}, which prevents us from gaining sufficient insight for optimal supercomputer operations.

For instance, one increasingly important use case for system-wide monitoring is to verify that resource usage is kept within acceptable margins and that power consumption levels meet specific power band requirements~\cite{sarood2014maximizing}. As soon as power exceeds a given bound, corrective actions must be taken by administrators in order to ensure system health, which is in turn monitored by a series of infrastructural and environmental sensors provided by facility management. On the other extreme, monitoring on a continuous basis, e.g., by sampling performance metrics from compute nodes, is critical to detect applications with potential bottlenecks. This, however, requires a completely different set of data sources~\cite{Guillen2014}, typically based on raw performance counter data integrated with derived application metrics. Even more so, ultimately, application behavior also has a direct impact on power consumption, making application data highly relevant to fine tune facility-wide power and energy management.

Further, in many scenarios, different applications or system components require concurrent access to the same data sources: as the employment of data analytics techniques to improve the efficiency of HPC systems becomes progressively common, more frameworks for fault tolerance~\cite{jones2012application}, runtime tuning and optimization~\cite{Eastep2017, periscope,Tapus:2002:AHT:762761.762771} and visualization, among others, require access to different types of data, such as performance counters, power meters, or even application- and runtime-level metrics. This leads to the necessity of a unified and controlled access to all data sources.

Given the above, a system monitoring solution must be \emph{holistic} (comprising of data sources of the facility, the system, the runtime, and the applications), \emph{thorough} (storing as many data points as available), and \emph{continuous} (analyzing and storing the data from all sensors at all times). In this paper, we present the \emph{Data Center Data Base} (\dcdb), a novel monitoring framework for HPC data centers, systems, and applications. It is designed following these requirements and hence addresses the complexity of managing new-generation installations for administrators and users alike.

Its key features are \emph{modularity}, which allows for easy integration and replacement in existing environments, and \emph{extensibility} through its plugin-based architecture, which enables addition of new data sources and suitability for a wide range of deployment requirements. \dcdb is also highly \emph{scalable}, thanks to its distributed and hierarchical architecture, and \emph{efficient}, due to its low-overhead implementation. The code is \emph{open-source}, and as such it can be freely customized according to the necessities of a specific data center. \dcdb is developed at the Leibniz Supercomputing Centre (LRZ) and is currently deployed on our production HPC systems. In particular we make the following contributions.

\begin{itemize}
\item We introduce the need and requirements for continuous monitoring in modern HPC facilities.
\item We implement \dcdb, a modular and extendable monitoring framework capable of combing sensor data from facilities all the way to applications.
\item We demonstrate its flexibility through a series of data collection agents covering a wide range of data sources.
\item We highlight its performance through a series of targeted experiments covering overheads of all individual components as well as overall scalability.
\item We show its applicability based on two case studies in the area of energy monitoring.
\end{itemize}

The remainder of the paper is structured as follows: Section~\ref{sections:challenges} briefly discusses the challenges that characterize HPC holistic monitoring, which we address in \dcdb. Section~\ref{sections:architecture} describes the design foundations of our framework and its architecture, and Section~\ref{sections:implementation} provides a detailed view of its implementation. Section~\ref{sections:interfaces} presents the available interfaces for data access in \dcdb. Section~\ref{sections:results} evaluates \dcdb's footprint on several production HPC systems, and Section~\ref{section:usecase} illustrates two real-world case studies using our framework. Finally, Section~\ref{sections:relatedwork} discusses related work on the topic of HPC monitoring, and Section~\ref{sections:conclusions} draws our conclusions and outlines future work.

\section{Monitoring Challenges}

\label{sections:challenges}
Establishing the necessary framework for holistic and continuous monitoring of large-scale HPC systems and their infrastructure is extremely challenging in many ways~\cite{ahlgren2018large, brandt2016large}.

\paragraph{(1)} One particular challenge is \emph{Scalability}: ``traditional'' sensor data (e.g., health status, temperature, power draw, network bandwidth) is comparatively easy and cheap to collect as it is typically acquired on a per-node basis and does not require very high readout frequencies. Even for large HPC systems, this type of data will only consist of a few thousand sensors which can still be handled by a monolithic and centralized monitoring solution. Application-related metrics (e.g., executed instructions, memory bandwidth, branch misses), however, typically need to be collected on a per-core basis and at high frequencies (i.e., 1Hz or higher). Consequently, they can easily add up to thousands of individual sensors per-compute node, resulting in millions of sensor readings per second, which creates bottlenecks particularly on large-scale HPC systems. Such vast amounts of sensor data can only be handled by a scalable and distributed monitoring solution.

\paragraph{(2)} The above also exposes a second challenge: \emph{Comparability}. Data from different sources is recorded at varying frequencies using varying units and exposing different metadata characteristics. We need mechanisms to translate data from different sensors, derive comparable metrics, and ultimately enable cross-source correlations.

\paragraph{(3)} A third challenge is \emph{Interference}: many metrics must be collected \emph{in-band} (i.e., on the compute nodes themselves) as opposed to \emph{out-of-band} (i.e., via a dedicated management network). For the former, monitoring solutions need to be highly-efficient in order not to interfere with running HPC applications, both in terms of \emph{Overhead} and \emph{Resource Footprint}, particularly \emph{Memory Footprint}. This is especially pressing when performing fine-granularity monitoring of several thousand metrics per-node, as discussed above, which could interfere with applications significantly.

\paragraph{(4)} Last, but not least, a challenge in monitoring is \emph{Extensibility}: very often, new devices (in hardware or software) need to be added to an already-running monitoring system, potentially requiring additional protocols and interfaces to access their data. Being able to easily add new ones, either by deploying existing implementations or developing them from scratch, is therefore important for production use of any monitoring solution.


\section{\dcdb Architecture}
\label{sections:architecture}

\begin{figure*}[t]
\centering
\includegraphics[width=0.90\textwidth,trim={0 0 0 0}, clip=true]{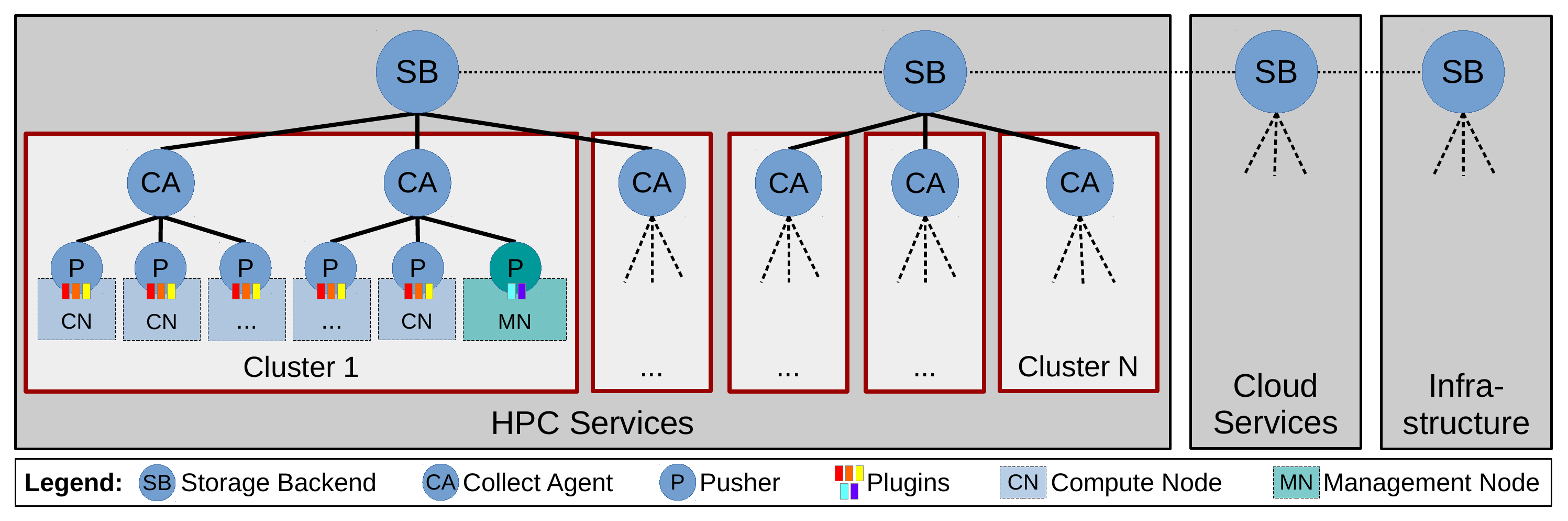}
\caption{A possible deployment scenario for \dcdb, highlighting its modular nature.}
\label{fig:dcdbArchHierarchy}
\end{figure*}

To address the challenges described in Section~\ref{sections:challenges}, \dcdb has been designed following a modular architecture, as depicted in Figure~\ref{fig:dcdbArchHierarchy}. We first describe the architecture in terms of its single components, and then explain the design principles on which it is based.

\subsection{Components of the Architecture}
\label{subsection:components}
\dcdb consists of three major classes of components, each with distinct roles: a set of data \emph{Pushers}, a set of \emph{Collect Agents} and a set of \emph{Storage Backends}. These components are distributed across the entire system and facility, which explicitly includes system nodes, facility management nodes and infrastructure components.

\paragraph{Transport Protocol.}
\dcdb employs the \emph{Message Queuing Telemetry Transport} (MQTT)~\cite{locke2010mq} protocol for the communication between Pushers and Collect Agents. We adopted MQTT for a variety of reasons: first, it is a lightweight protocol that was introduced specifically for the exchange of telemetry data, which makes it an ideal fit for sensor monitoring. Moreover, it has a generic design and is well-established, enabling \dcdb to be compatible with a vast amount of available tools for data transmission, processing and visualization that support MQTT. Similarly, several implementations of MQTT exist for a large variety of platforms and architectures; in this regard, by exploiting the modular design of \dcdb, developers can choose to swap one Pusher against another as long as it also employs MQTT for transmitting data.

The protocol is based on a publish/subscribe model in which senders publish their messages under a certain topic to which potential receivers can subscribe. MQTT topics are essentially strings that describe the content of each message and are organized similarly to file system path names, i.e., they implicitly define a hierarchy. We leverage this feature in \dcdb by associating a unique MQTT topic to each sensor, thus defining a sensor hierarchy. The individual hierarchy levels can be defined by the user, but we commonly specify them reflecting the location of the monitored entities (e.g., a hierarchy would comprise levels associated to rooms, systems, racks, chassis, nodes, and CPUs). Defining an appropriate hierarchy for sensors is fundamental in a holistic monitoring tool such as \dcdb: since the amount of available sensors can potentially become enormous due to system size and to the monitoring requirements of different parties (e.g., developers, vendors, operators), enabling separation of the sensor space greatly improves the navigability, and in turn, the usability of the \dcdb framework itself.

\paragraph{Pusher.}
A Pusher component is responsible for collecting monitoring data and is designed to either run on a compute node of an HPC system to collect in-band data or on a management or facility server to gather out-of-band data. The plugins for the actual data acquisition are implemented as dynamic libraries, which can be loaded at initialization time as well as at runtime. We currently provide ten different Pusher plugins, supporting in-band application performance metrics (Perfevents~\cite{weaver2013linux}), server-side sensors and metrics (ProcFS\footnote{\url{http://man7.org/linux/man-pages/man5/proc.5.html}} and SysFS), I/O metrics (GPFS and Omnipath), out-of-band sensors of IT components (IPMI~\cite{intel2010ipmi} and SNMP~\cite{rfc1157}), RESTful APIs, and building management systems (BACnet~\cite{ashrae2010bacnet}). A Pusher's data collection capabilities are only limited by the available plugins and their supported protocols and data sources, and it is therefore adaptable to a wide variety of use cases.

\paragraph{Collect Agent.}
A Collect Agent is responsible for receiving the sensor readings from a set of associated Pushers and writing them to a Storage Backend. For that purpose, it assumes the role of an MQTT broker that manages the publish/subscribe semantics of the MQTT protocol: Pushers publish the readings of individual sensors under their specific topics and Collect Agents forward them to the potential subscribers of those topics. In the current design of \dcdb, the Storage Backend is the only subscriber that subscribes to all MQTT topics. However, it is possible that additional subscribers may want to receive certain sensor readings as well for other purposes, for example for on-the-fly analysis of data or online tuning.

\paragraph{Storage Backend.}
By its nature, monitoring data is time series data that is typically acquired and consumed in bulk: it is streamed into the database and retrieved for longer time spans, and not for single points in time. Logically, the data points for a sensor are organized as a tuple of \emph{<sensor, timestamp, reading>}. These properties make monitoring data a perfect fit for noSQL databases in general and wide-column stores in particular, due to their high ingest and retrieval performance for this kind of streaming data.

The current implementation of \dcdb leverages \emph{Apache Cassandra}~\cite{wang2012nosql} for Storage Backends, but due to its modularity it could easily be swapped for a different database such as \emph{InfluxDB}\footnote{\url{https://www.influxdata.com/}}, \emph{KairosDB}\footnote{\url{https://kairosdb.github.io/}}, or \emph{OpenTSDB}\footnote{\url{http://opentsdb.net}}. We chose Cassandra due to its data distribution mechanism that allows us to distribute a single database over multiple server nodes, or Storage Backends, either for redundancy, scalability, or both. This feature works in synergy with the hierarchical and distributed architecture of \dcdb and effectively allow us to scale our system to arbitrary size.

\subsection{Design Principles}
Here we introduce the main design principles and requirements driving the architecture of \dcdb (\emph{Data Flow} and \emph{Components}), alongside the concepts around which the framework itself and its interfaces are built (\emph{Sensors} and \emph{Virtual Sensors}).

\paragraph{Data Flow.}
The data collection mechanism follows the push principle: instead of a central server that pulls data from monitored entities, a distributed set of Pushers close to the data sources acquire data and push it to the Collect Agents. Collect Agents receive monitoring data from their associated Pushers and forward it to their respective Storage Backend for persistent storage, which (in combination) form the overall distributed \dcdb data storage. 

Communication between components is performed via established protocols and well-defined APIs such that each component could be easily swapped for a different implementation, leveraged for other purposes or integrated into existing environments. 

\paragraph{Components.}
The modular design of \dcdb facilitates its scalability as all components are designed to be distributed and are hierarchically-organized: in a typical setup, there will be a large number of Pushers (hundreds or thousands), many Collect Agents (in the order of dozens), and one or more Storage Backends. Depending on the system's size and the number of metrics to be monitored, the number of Pushers, Collect Agents and Storage Backends can be scaled to handle the load and to sustain the required ingest rates. In terms of extensibility, Pushers provide a flexible plugin-based interface that allows for easily adding new and different data sources via various protocols and interfaces. \dcdb currently provides plugins for the most commonly-used protocols, but additional plugins can be implemented with low coding effort.

\paragraph{Sensors.}
In the context of \dcdb, each data point of a monitored entity is called a \emph{sensor}. This could be a physical sensor measuring temperature, humidity or power, but can also be any other source of monitoring data such as a performance counter event of a CPU, the measured bandwidth of a network link or the energy meter of a power distribution unit (PDU). Each sensor's data consists of a time series, in which readings are represented by a timestamp and a numerical value. This format is enforced across \dcdb, ensuring consistency and uniformity of sensor data.

\paragraph{Virtual Sensors.}
\dcdb supports the definition of \emph{virtual sensors}, which supply a layer of abstraction over raw sensor data and can be used to provide derived or converted metrics. They are generated according to user-specified arithmetic expressions of arbitrary length, whose operands may either be sensors or virtual sensors themselves. This can be used, for instance, to aggregate data from several sources in order to gain insight about the status of a system as a whole (e.g., aggregating the power sensors of individual compute nodes in an HPC system), or to calculate key performance indicators such as the Power Usage Effectiveness (PUE) from physical units measured by sensors. Virtual sensors can be used like normal sensors and are evaluated lazily, i.e., they are only computed upon a query and only for the queried period of time. As queries to virtual sensors may potentially be expensive (in terms of computation as well as I/O), results of previous queries are written back to a Storage Backend so they can be re-used later. The units of the underlying physical sensors are converted automatically and we account for different sampling frequencies by linear interpolation.
\section{\dcdb Implementation}
\label{sections:implementation}
\begin{figure}[t]
\centering
\includegraphics[width=0.47\textwidth,trim={0 0 0 0}, clip=true]{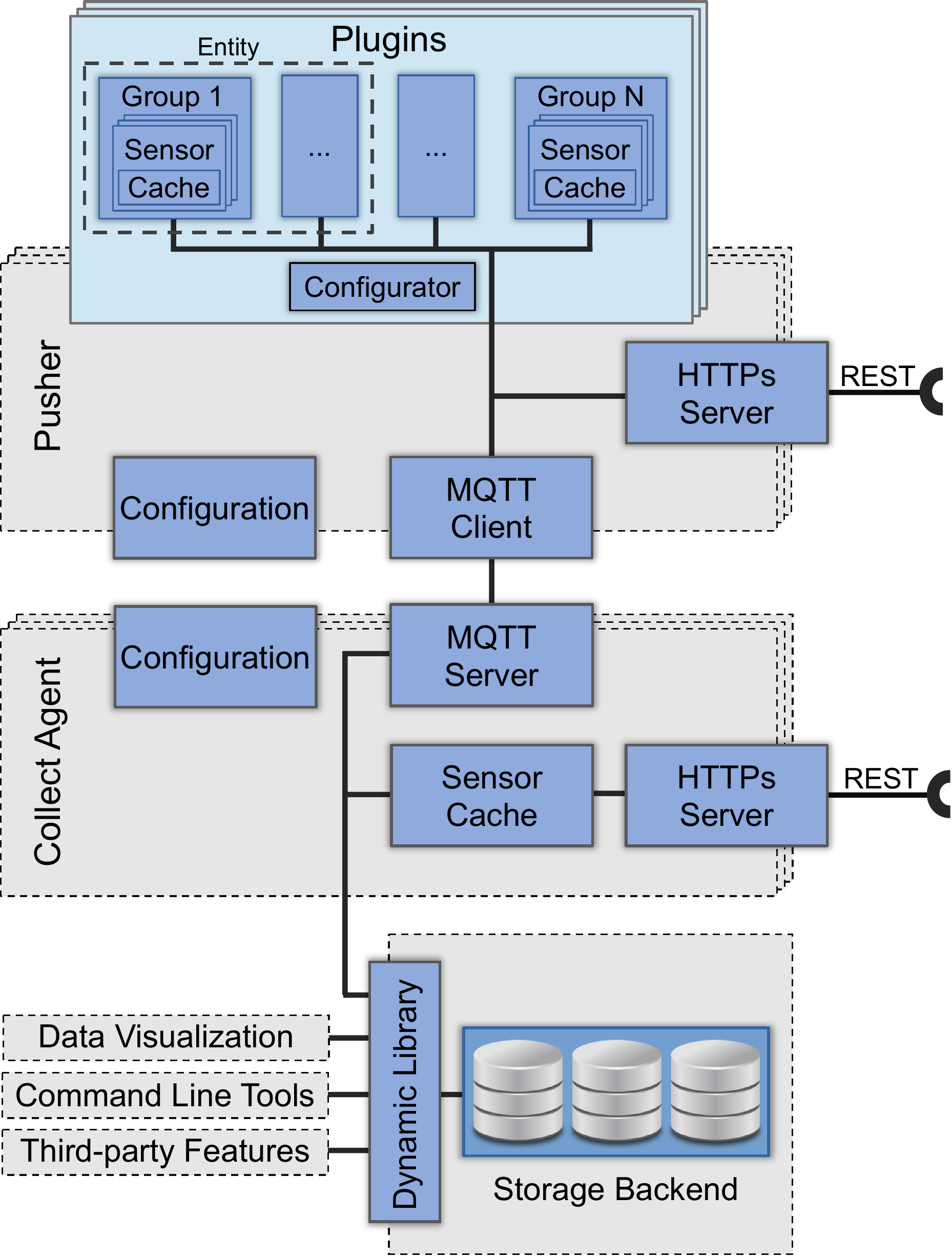}
\caption{Architectural overview of \dcdb highlighting the data flow from sensors (top) to storage (bottom). Note how entities can be optionally used to combine sensor groups and allow them to share resources.}
\label{fig:dcdbArchVert2}
\end{figure}

\dcdb is written in C++11 and is freely available under the GNU GPL license via GitLab\footnote{https://dcdb.it}.
In this section we provide greater detail on the implementation specifics on \dcdb's core components, which are also represented in Figure~\ref{fig:dcdbArchVert2}. 

\subsection{Pusher Structure}
A Pusher comprises a set of \emph{Plugins}, an \emph{MQTT Client}, an \emph{HTTPs Server}, and a \emph{Configuration} component. The latter is responsible for configuring the Pusher at start-up and instantiating the required plugins. This process is controlled by a set of configuration files that define the data sources for each plugin as well as the global Pusher configuration. The configuration files use an intuitive \emph{property tree} format, which is documented on the \dcdb code repository with several examples. The HTTPs Server provides a RESTful API to facilitate configuration tasks and to access the sensor caches in the plugins (see Section~\ref{sections:restapi} for details). The MQTT Client component periodically extracts the data from the sensors in each plugin and pushes it to the associated Collect Agent. It relies on the \emph{Mosquitto} library~\cite{light2017mosquitto} for MQTT communication, which proved to be the most suitable for our purposes in terms of scalability, stability and resource footprint. The key components of a Pusher, however, are the plugins that perform the actual data acquisition and consist of up to four logical components:

\paragraph{Sensors} The most basic unit for data collection. A sensor represents a single data source that cannot be divided any further. It may represent, e.g., the L1 cache misses of a CPU core or the power consumption of a device, which are sampled and collected as a numerical time series. A sensor always has to be part of a \emph{group}.
\paragraph{Groups} The next aggregation level combining multiple sensors. All sensors that belong to one group share the same sampling interval and are always read collectively at the same point in time. Groups are intended to tie together logically-related sensors, such as all power outlets of one power delivery unit and cache-related performance counters of a CPU core.
\paragraph{Entities} An optional hierarchy level to aggregate groups or to provide additional functionality to them. For example, for a plugin reading data from a remote server (e.g., via IPMI or SNMP), a host entity may be used by all groups reading from the same host for communication with it.
\paragraph{Configurator} The component responsible for reading the configuration file of a plugin and instantiating all components for data collection. It provides the interface between the Pusher and one of its plugins, and gives access to its entities, groups and sensors.

\bigbreak
System administrators deploying \dcdb are encouraged to extend Pushers according to their own needs by implementing plugins for new data sources. To simplify the process of implementing such plugins \dcdb provides a series of \emph{generator} scripts. They create all files required for a new plugin and fill them with code skeletons to connect to the plugin interface. Comment blocks point to all locations where custom code has to be provided, greatly reducing the effort required to implement a new plugin.

Sensor read intervals are not only synchronized within groups, but also across plugins and even Pushers by means of the \emph{Network Time Protocol} (NTP)~\cite{mills1991internet}. Moreover, \dcdb's push-based monitoring approach allows for more precise timings compared to pull-based monitoring, especially at fine-grained (i.e., sub-second) sampling intervals. This allows for easily correlating different sensors without having to interpolate readings to account for different readout timestamps. Additionally, this minimizes jitter on compute nodes of HPC systems as parallel applications running on multiple nodes will be interrupted at the same time and hence less load imbalance will be introduced~\cite{ferreira2008characterizing}. Although the data collection intervals of multiple Pushers are synchronized, these will send their data at different points in time in order not to overwhelm the network, Collect Agents and Storage Backends.

\subsection{Collect Agent as Data Broker}
Collect Agents are built on top of a custom MQTT implementation that only provides a subset of features necessary for their tasks. In particular, it only supports the publish interface of the MQTT standard, but not the subscribe interface. As Storage Backends are currently the only consumers of sensor readings, this avoids additional overhead for filtering MQTT topics. 

Upon retrieval of an MQTT message, a Collect Agent parses the topic of the message and translates it into a unique numerical Sensor ID (SID) that is used as the key to store a sensor's reading in a Storage Backend. There is a 1:1 mapping of topics to SIDs which maintains the hierarchical organization of MQTT topics: each topic is split into its hierarchical components and each such component is mapped to a numeric value that is stored in a particular bit field of the 128-bit SID.

\subsection{Cassandra Storage Backends}
As mentioned in Section~\ref{sections:architecture}, we picked Apache Cassandra for our Storage Backends as it fits well with the semantics of monitoring data and because its distributed approach allows for the scalability required of a holistic monitoring framework.

As Cassandra may be distributed across multiple servers (a ``cluster'' in Cassandra terminology), any of those servers may be used to insert or query data. The distribution of data within the cluster can be controlled via partition keys and a partitioning algorithm. We exploit this feature by leveraging the hierarchical SIDs as partition keys for Cassandra: using a partitioning algorithm that maps a sub-tree in the sensor hierarchy to a particular database server allows for storing a sensor's reading on the nearest server and thus to avoid network traffic. The same logic is applied for queries to minimize network traffic between the database servers by directing them directly to the respective server. This logic is implemented in lib\dcdb (see Section~\ref{sections:libdcdb}) and is fully transparent to Collect Agents as well as to the users requesting access to sensor data.

\section{\dcdb Interfaces}
\label{sections:interfaces}
\dcdb provides several interfaces to access the stored monitoring data. Users and administrators of an HPC system can perform this task with the support of a specific dynamic library (lib\dcdb), with command line tools that are built on top of this library and via RESTful APIs. Furthermore, data can be converted to be analyzed using the Grafana visualization tool of GrafanaLabs\footnote{\url{https://grafana.com/}}.

\subsection{lib\dcdb}
\label{sections:libdcdb}
All accesses to Storage Backends are performed via a well-defined API that is independent from the underlying database implementation. While we use Apache Cassandra in our current Storage Backends, this abstraction allows for easily swapping it against a different database solution without any changes in the upstream components. Currently, the API has only been implemented in a C++ library, lib\dcdb, but other bindings could  easily be implemented as well. Additionally, the C++ library can also be used in Python scripts and hence covers a wide range of use cases.

\subsection{Command Line Tools}
\dcdb offers a series of command line tools that leverage lib\dcdb for access to Storage Backends. Among these, the \emph{config} tool allows administrators to perform basic database management tasks (e.g., deleting old data or compacting) as well as configuring the properties of sensors such as units and scaling factors or defining virtual sensors. The \emph{query} tool then allows users to obtain sensor data for a specified time period in CSV format or perform basic analysis tasks on the data such as integrals or derivatives. A series of secondary tools offers utility features, like a \emph{csvimport} tool to import CSV data into Storage Backends.

\subsection{RESTful APIs}
\label{sections:restapi}
Pushers and Collect Agents further support data retrieval through RESTful APIs. In a Pusher this provides an interface to retrieve the current configuration (e.g., of plugins or sensors) and allows for starting and stopping individual plugins. This can be useful, for example, to avoid conflicts with user software accessing the same data source, or to enable additional data sources for individual applications. Additionally, one can modify a plugin's configuration file at runtime and trigger a reload of the configuration, which allows a seamless re-configuration without interrupting the Pusher. Further, the RESTful API also provides access to a \emph{sensor cache} that stores the latest readings of all sensors. It is configurable in size and can be used by other processes (either on the same machine or via the network) to easily read all kinds of sensors via a common interface from user space. 

Analogous to Pushers, Collect Agents provide a sensor cache that can be queried via the same RESTful API and that gives access to the most recent readings of all Pushers connected to them. This can be used, for example, to feed all readings into another (legacy) monitoring framework without having to deal with the protocols of various sensors.

\subsection{Visualization of Data}
\begin{figure}[b]
\centering
\includegraphics[width=0.45\textwidth,trim={0 0 0 0}, clip=true]{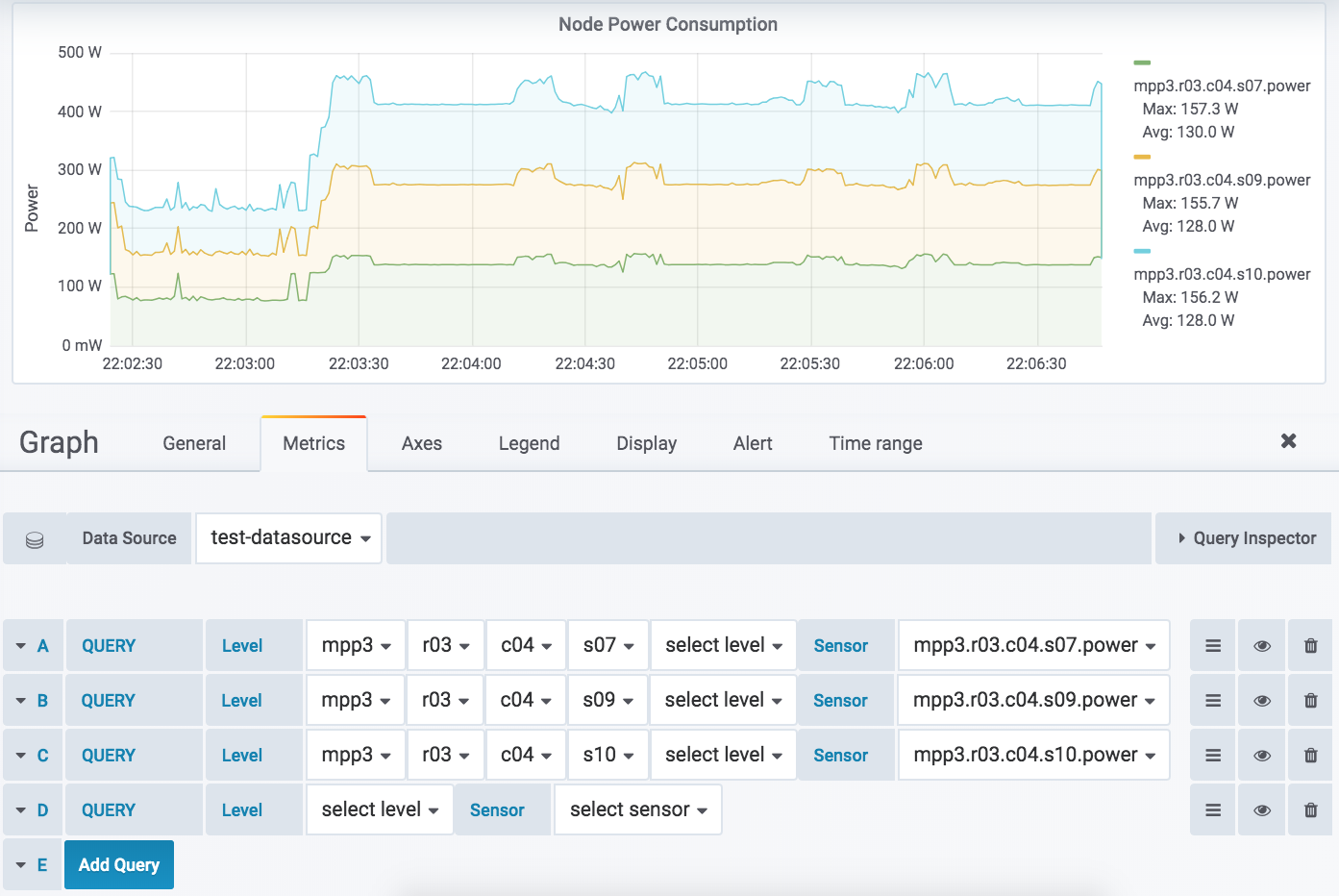}
\caption{The Grafana data source panel for \dcdb.}
\label{fig:grafana}
\end{figure}

\dcdb leverages Grafana for the visualization of monitoring data. Among its many benefits, Grafana suits our needs primarily because a) it provides a comprehensive set of visualization options (e.g., graphs, heatmaps, histograms or tables); b) it allows users to define alerts and receive associated notifications via multiple channels; c) it is designed following an extensible architecture, allowing to develop dedicated plugins; d) it has a strong user and development community; and e) is completely open-source.

However, although Grafana supports several database implementations, it does not provide any plugin for Apache Cassandra. We therefore develop our own plugin that leverages lib\dcdb. In addition to retrieving data from Cassandra, it is also designed to profit from current and future features offered by \dcdb.

To this day, a missing feature in Grafana and in all of its plugins for different databases is the possibility to build hierarchical queries and to select metrics at a specific level of the hierarchy. This functionality is useful in HPC or data center environments, where a system administrator can browse different hierarchical levels of a system (e.g., a rack, a chassis, or a server) and query data from sensors available at that level. This becomes particularly beneficial if the monitored system comprises a very large number of sensors (potentially in the order of millions for leadership-class HPC systems). As \dcdb employs such a hierarchy on all sensors (see Section~\ref{sections:architecture}), our data source plugin also exposes it in Grafana.

Figure~\ref{fig:grafana} illustrates a visualization example of this feature, specifically plotting the power consumption of three different nodes on one of our production systems at LRZ. As depicted, users can query sensors at a specific hierarchical level by navigating through the hierarchy with the support of multiple drop-down menus. The visualization of data may further benefit from convenient features such as stacking of time series data or comprehensive formatting of axes and legends (e.g., displaying useful information like current average or maximum values of the plotted metrics).

\section{Performance and Scalability}
\label{sections:results}

\begin{table*}[t]
\caption{The architectures of three production environments at LRZ, together with the corresponding per-node Pusher configurations and average overhead values as measured against the HPL benchmark.}
\label{table:productionconfig}
\begin{tabular}{c|c|c|c|c||c|c|c}
\textbf{HPC System} & \textbf{Nodes} & \textbf{CPU} & \textbf{Memory} & \textbf{Interconnect} & \textbf{Plugins} & \textbf{Sensors} & \textbf{Overhead} \\ \hline 
SuperMUC-NG         & \makecell{6480 \\ Skylake} & \makecell{Intel Xeon Platinum 8174 \\ 2 cpus x 24 cores x 2 threads}  & 96GB & \makecell{Intel \\ OmniPath} & \makecell{Perfevents, ProcFS, \\ SysFS, OPA}       & 2477 & 1.77\%             \\ \hline
CooLMUC-2         & \makecell{384 \\ Haswell} & \makecell{Intel Xeon E5-2697 v3 \\ 2 cpus x 14 cores} & 64GB & \makecell{Mellanox \\ Infiniband} & \makecell{Perfevents, ProcFS, \\ SysFS} & 750 & 0.69\%              \\ \hline
CooLMUC-3 & \makecell{148 \\ KNL} & \makecell{Intel Xeon Phi 7210-F \\ 64 cores x 4 threads} & \makecell{96GB \\ 16GB HBM} & \makecell{Intel \\ OmniPath} & \makecell{Perfevents, ProcFS, \\ SysFS, OPA}  & 3176    & 4.14\%         
\end{tabular}
\end{table*}

In this section we discuss the performance of \dcdb from different viewpoints. We start by evaluating Pushers in production and test configurations, and then proceed by analyzing Collect Agents. Our purpose is to quantify the performance of our framework as a whole: by quantifying the impact of a Pusher on running applications we assess its overhead, whereas by analyzing its resource usage we characterize its footprint and scalability. The same approach is applied to a Collect Agent to characterize its scalability at various data rates, and thus to prove the suitability of \dcdb for extreme-scale HPC installations.

\subsection{Experimental Setup}
\label{section:expsetup}

\paragraph{Reference Applications.} In order to estimate how \dcdb impacts running applications, we performed tests by executing instances of well-known benchmarks on multiple node architectures. We first present a series of tests performed in a production environment (Section \ref{section:productionperf}), where we aim to characterize the impact of \dcdb on real applications sensitive to network and memory bandwidth, and on MPI communication. For this reason, we employ a selection of MPI benchmarks from the CORAL-2 suite\footnote{\url{https://asc.llnl.gov/coral-2-benchmarks}}, namely \emph{Quicksilver} \cite{richards2017quicksilver}, \emph{LAMMPS} \cite{plimpton1995fast}, \emph{AMG} \cite{yang2002boomeramg} and \emph{Kripke} \cite{kunen2015kripke}. These four benchmarks cover a large portion of the behavior spectrum of HPC applications, and results obtained with these can be considered representative of real work loads.

Later on, we characterize the impact of \dcdb on computational resources and its scalability with various test configurations (Section~\ref{section:testperf}), stressing the communication and sampling subsystems. In this part, we focus on using the shared-memory version of the \emph{High-Performance Linpack} (HPL) benchmark~\cite{dongarra2003linpack}, supplied with the Intel MKL library\footnote{\url{https://software.intel.com/en-us/articles/intel-mkl-benchmarks-suite}}. Being a compute-bound application, tests performed against HPL give us insights on the behavior of \dcdb in a worst-case scenario. 

\paragraph{Test Architectures.} For the Pusher-related part of this analysis we use nodes from three different production HPC systems at LRZ, each with a different architecture (see Table~\ref{table:productionconfig}). The \emph{Skylake} and \emph{Haswell} CPUs provide strong single-thread performance, whereas the \emph{Knights Landing} CPU with its large number of (SMT-) cores is comparatively weak in this regard. A Collect Agent was running on a dedicated database node, equipped with two Intel E5-2650 v2 CPUs, 64GBs of RAM and a 240GB Viking Tech SSD drive.

\paragraph{Configuration Parameters.}
All benchmarks are configured to instantiate one MPI process (in case of MPI codes) per node, and use as many OpenMP threads as physical CPU cores that are available. Pushers are configured to use two sampling threads and a sensor cache size of two minutes (see Section~\ref{sections:restapi}).

\paragraph{Evaluation Metrics.} Each experiment involving benchmark runs was repeated 10 times to ensure statistical significance. To account for outliers and performance fluctuations, we use median runtimes. We then use the following metrics to evaluate the performance of the \dcdb components:

\begin{itemize}
    \item \textbf{Overhead} is defined as the fraction of time an application spends in excess compared to running without \dcdb due to interference from Pushers. We obtain it by comparing the \emph{reference} execution time (\(T_r\)) of an application run against the one observed when a Pusher is run (\(T_p\)), and we compute it as \(O = (T_p - T_r) / T_r \). The overhead helps quantify the impact of a component on the system performance. We compute it in terms of the runtime impact on reference applications, so as to obtain scalable and reproducible experiments. Our evaluation process does not include system throughput among the selected metrics as it would significantly depend on the underlying workload, and as such would require a large-scale dedicated environment. Nevertheless, based on our experience on our production systems, we do expect low runtime overhead to directly translate into low throughput change.
    \item \textbf{CPU Load} is defined as the percentage of active CPU time spent by a process against its total runtime, as measured by the Linux \emph{ps} command; this metric characterizes a Pusher's and a Collect Agent's performance when used in a out-of-band context, with no overhead concerns.
    \item \textbf{Memory Usage} of a process is quantified by \emph{ps}. It helps characterize the impact of different monitoring configurations in Pushers and Collect Agents.
\end{itemize}

\subsection{Pusher Performance}
\label{section:pusherperf}
First, we present the performance of Pushers in terms of computational overhead against work loads running in an HPC system. 

\subsubsection{Overhead in a Production Configuration}
\label{section:productionperf}

\begin{figure}[b]
\centering
\includegraphics[width=0.45\textwidth,trim={0 0 0 0}, clip=true]{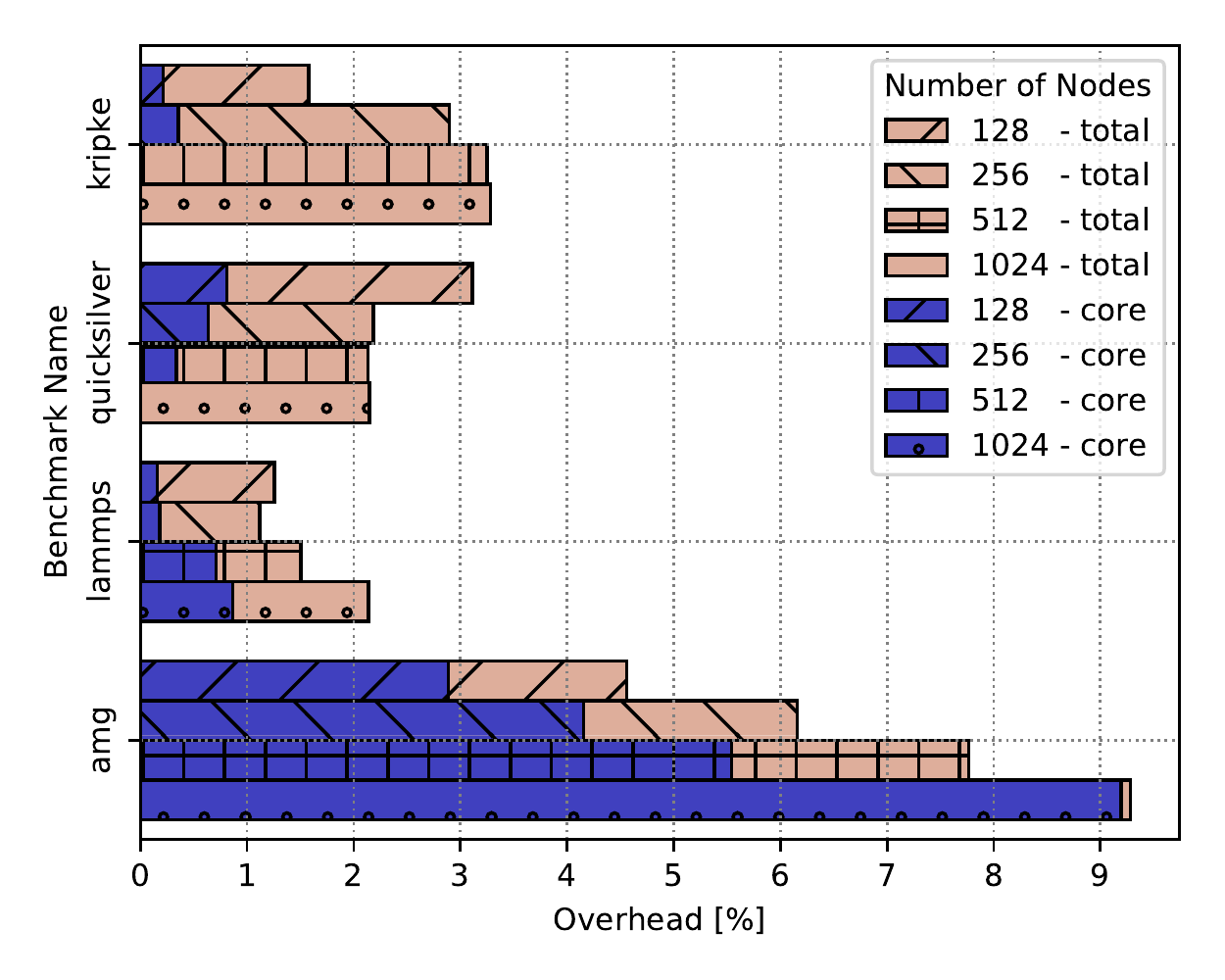}
\caption{Pusher overhead on CORAL-2 MPI benchmarks using production (total) and test (core) setups on the SuperMUC-NG system.}
\label{results:mpioverhead}
\end{figure}

\begin{figure*}[t]
 \centering
 \captionsetup[subfigure]{}
    \subfloat[Overhead on the Skylake architecture.]{
    \includegraphics[width=0.3\textwidth,trim={0 0 100 0}, clip=true]{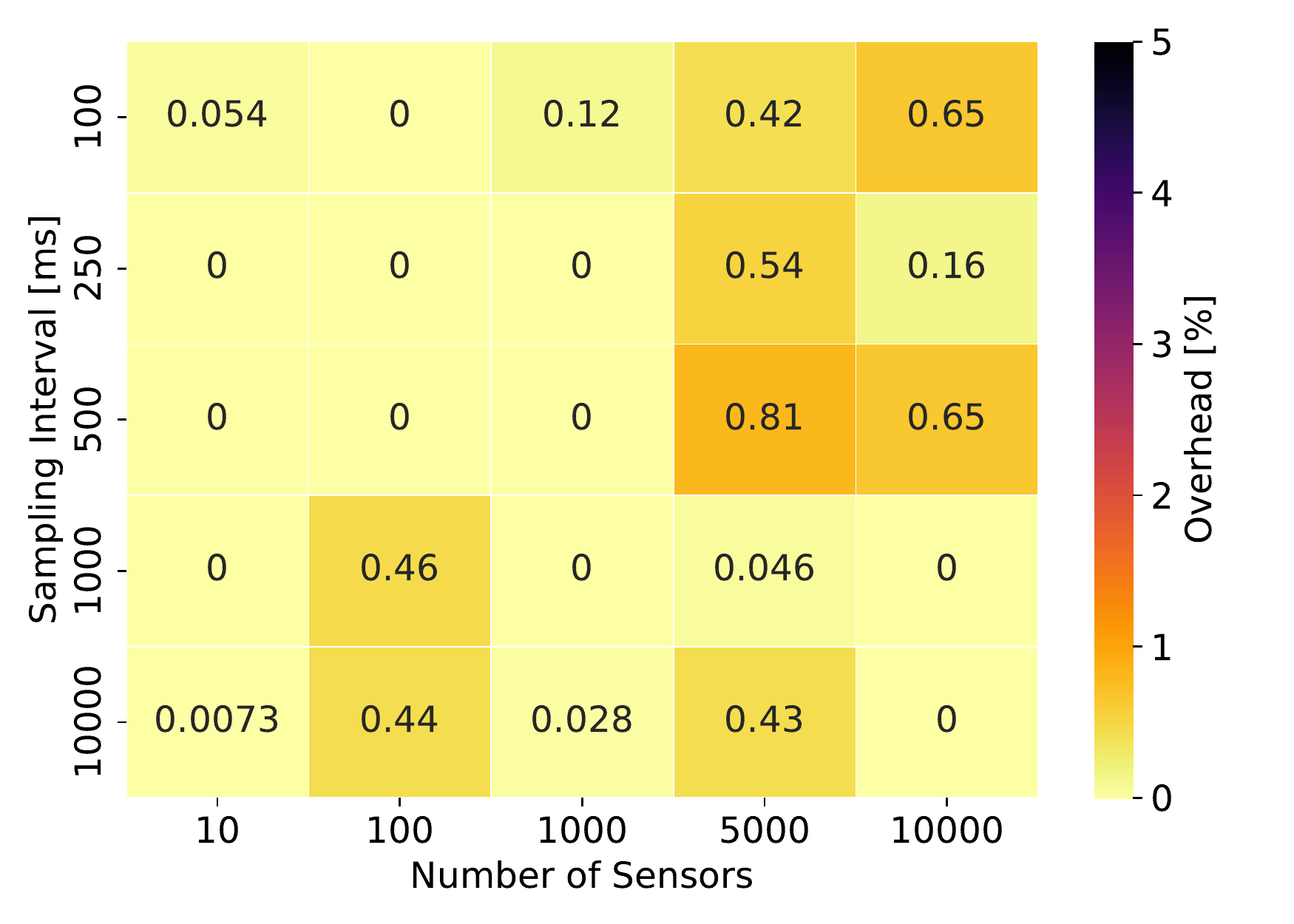}
  	}
  \hfill
  \subfloat[Overhead on the Haswell architecture.]{
    \includegraphics[width=0.2825\textwidth,trim={25 0 100 0}, clip=true]{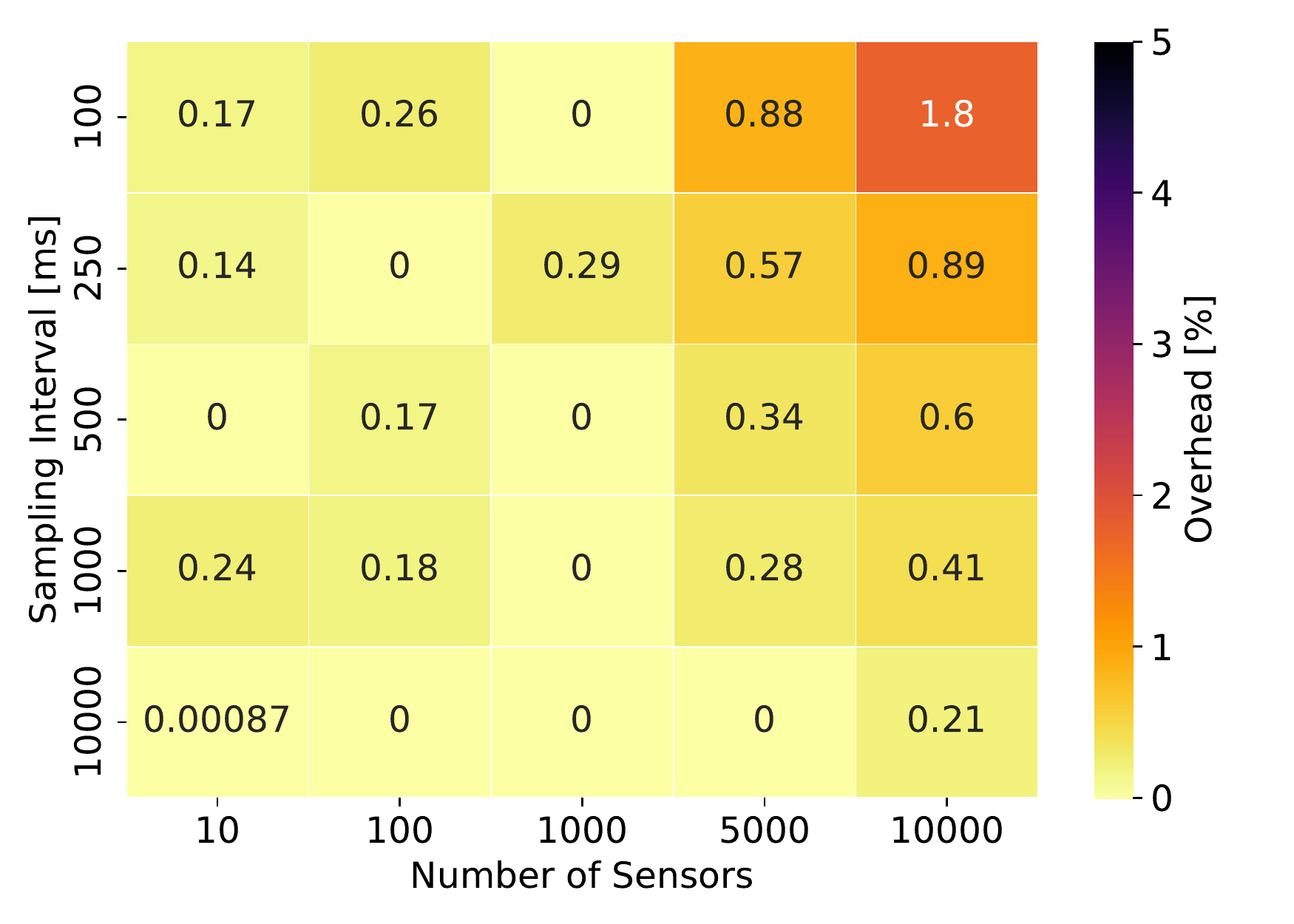}
  }
  \hfill
  \subfloat[Overhead on the Knights Landing architecture.]{
    \includegraphics[width=0.359\textwidth,trim={25 0 0 0}, clip=true]{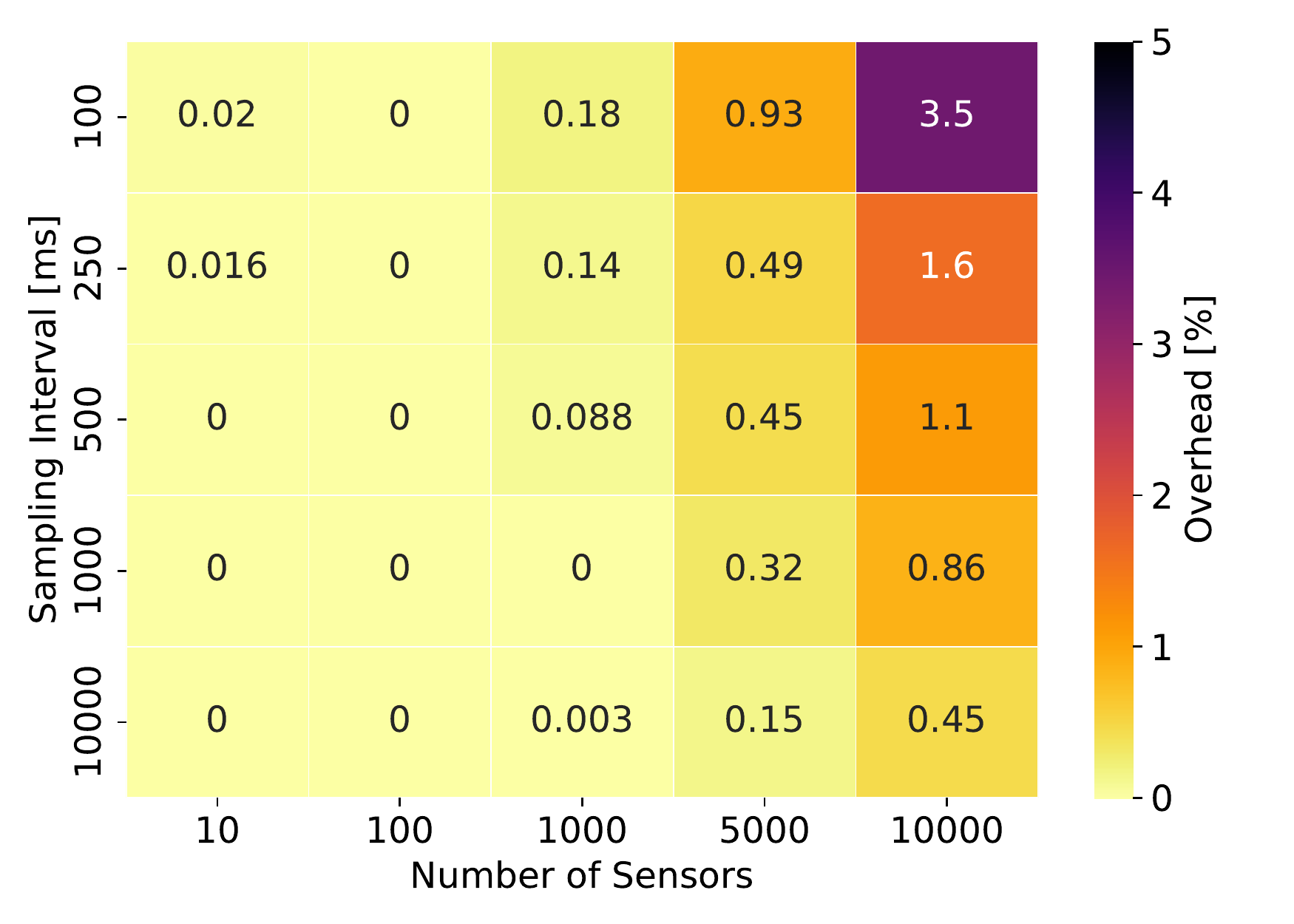}
  }
  \caption{Heatmaps of the Pushers' overhead at various sampling intervals and sensor numbers, on three different HPC architectures, against the HPL benchmark.}
  \label{results:heatmaps}
 \end{figure*}

To assess a Pusher's performance in typical production environments, we are using the actual configurations deployed on our systems as described in Table~\ref{table:productionconfig}. In all configurations, the ProcFS plugin collects data from the \emph{meminfo}, \emph{vmstat} and \emph{procstat} files, whereas we use SysFS to sample various temperature and energy sensors. we use Perfevents to sample performance counters on CPU cores, and finally OPA to measure network-related metrics. For this use case, we only deploy plugins that perform in-band measurement with a sampling interval of 1 second; out-of-band measurement would be performed on separate machines and hence does not incur overhead on running applications. Additionally, in some experiments we only deploy the \emph{tester} plugin, which can generate an arbitrary number of sensors with negligible overhead. This allows us to isolate the overhead of the various monitoring backends (e.g., IPMI or perfevents) from that of the Pusher, which is mostly communication-related. 

We measured the overhead against the CORAL-2 MPI benchmarks with different node counts on our Skylake-based SuperMUC-NG system, using a weak scaling approach. We present our results in Figure~\ref{results:mpioverhead}. The experiment was performed twice: once with the Pusher configuration presented in Table~\ref{table:productionconfig} (labeled \emph{total}), and once using a configuration with the same number of sensors, produced with the tester plugin (labeled \emph{core}). The overhead for LAMMPS, Quicksilver and Kripke is low and never goes above 3\%. Moreover, when scaling the number of nodes, the overhead increase is minimal. The AMG benchmark represents an exception, showing a linear increase with respect to the node count, and peaking at 9\% with 1024 nodes: this application is notorious for using many small MPI messages and fine-granular synchronization. Consequently, it is extremely sensitive to network interference. This is also confirmed by the experiments with the tester plugin: LAMMPS, Quicksilver and Kripke are affected to a very limited extent by the Pushers' network interference, whereas in AMG this causes most of the total overhead. Moreover, we observe the best performance for AMG when the Pushers were configured to send sensor data to a Collect Agent in regular bursts twice per minute, reducing network interference. The remaining benchmarks, on the other hand, perform better when the Pushers' data is sent out continuously in a non-bursty manner. This type of interference can be avoided by using a separate network interface (e.g., for management) to transmit data from compute nodes.

We present overhead results against single-node HPL runs for all three architectures in Table~\ref{table:productionconfig}. In most configurations the overhead is low despite the large number of sensors being pushed each second. The worst performer is the Knights Landing architecture: this was expected due to its weak single-thread performance and the much larger number of collected sensors than in the Skylake and Haswell configurations, which is due to the large number of SMT cores on this architecture. Average memory usage ranges between 25MB (Haswell) and 72MB (Knights Landing), whereas average per-core CPU load ranges between 1\% (Haswell) and 9\% (Knights Landing).

\subsubsection{Overhead in a Test Configuration}
\label{section:testperf}

\begin{figure}[t]
\centering
\captionsetup[subfigure]{}
    \subfloat[Average per-core CPU load.]{
    \includegraphics[width=0.45\textwidth,trim={0 0 0 0}, clip=true]{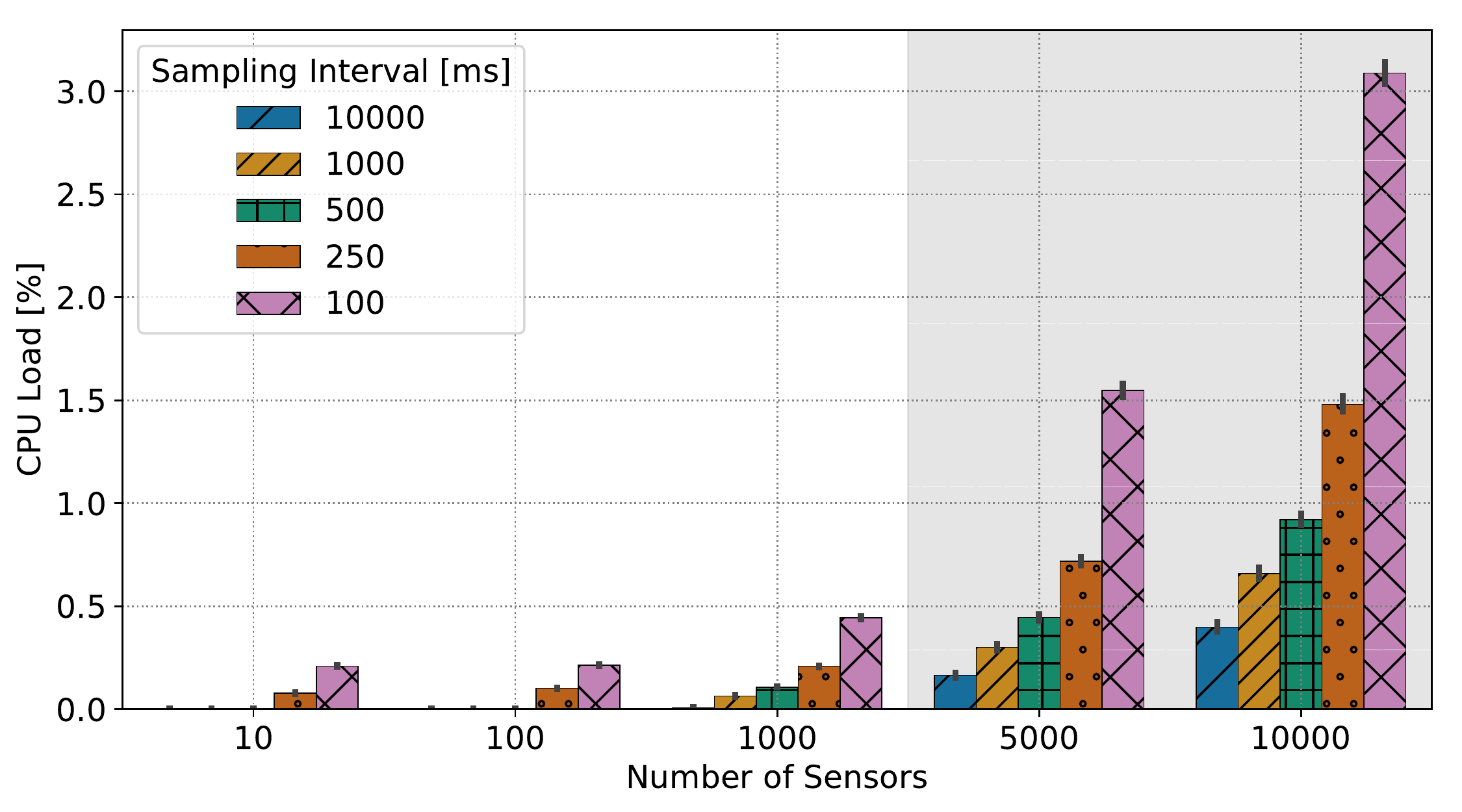}
  	}
  \\	
  \subfloat[Average memory usage.]{
    \includegraphics[width=0.45\textwidth,trim={0 0 0 0}, clip=true]{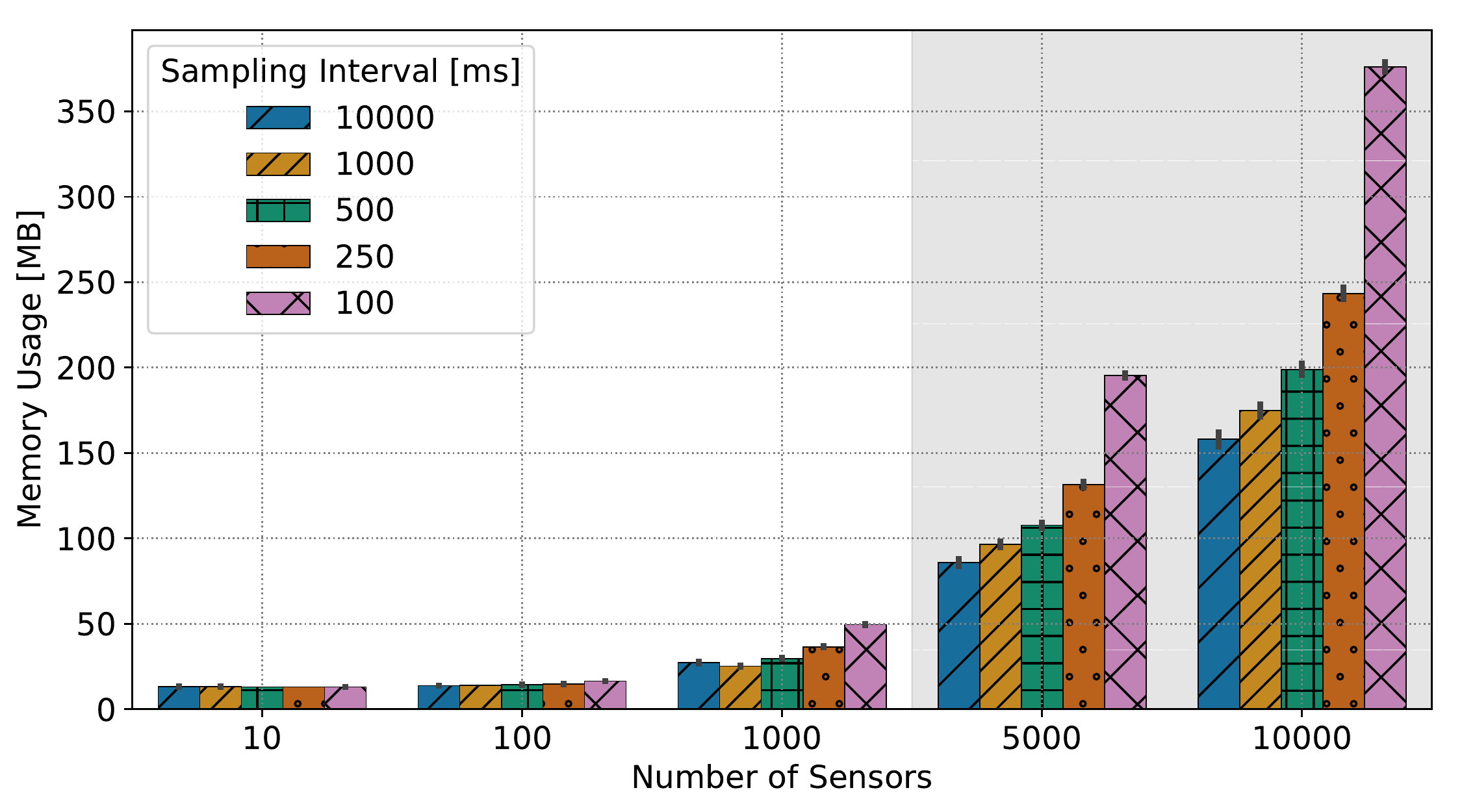}
  }
  \caption{Average Pusher CPU load and memory usage on SuperMUC-NG nodes. The white area highlights typical configurations for production HPC environments.}
  \label{results:resourceusage}
\end{figure}

In this second part we estimate the scalability of the Pusher's core, once again using the tester plugin. We analyze a total of 25 configurations, which differ in terms of sampling intervals and number of sensors. Figure~\ref{results:heatmaps} depicts the results in terms of overhead against single-node HPL runs, for each of the three analyzed node architectures. In the plot, a value of 0 denotes no overhead, meaning that the median runtime when using a Pusher in the experiment was equal or less than the reference median runtime. In all cases the computational overhead is low, and in all configurations with 1,000 sensors or less, which are typical for production environments, it is below 1\%. Even when pushing 100,000 sensor readings per second (10,000 sensors sampled every 100ms) overhead remains acceptable for all platforms. The Skylake node architecture, in particular, is unaffected by the various Pusher configurations and shows consistent overhead values. Haswell and Knights Landing show clearer gradients with increasing overhead in the most intensive configurations, with the latter of the two exhibiting the worse results due to its weak single-thread performance. \dcdb is thus usable in scenarios with large numbers of sensors and at high sampling rates.

In Figure~\ref{results:resourceusage} we show results in terms of average per-core CPU load and memory usage for each configuration. We show results only for the Skylake-based SuperMUC-NG nodes, since all node types scale similarly. Memory usage is dependent on both the sampling interval and number of sensors, as these will result in different sensor cache sizes. In the most intensive configuration with 100,000 sensor readings per second, memory usage averages at 350MB, and is well below 50MB for typical production configurations, that have 1,000 sensors or less. It can be further reduced by tuning the size of sensor caches. CPU load peaks at 3\% in the most intensive configuration, proving that there is ample room for much more intensive and fine-granularity configurations, in environments in which computational overhead is not a concern, such as when performing out-of-band monitoring.

\subsection{Performance Scaling Modeling}

\begin{figure}[t]
 \centering
 \includegraphics[width=0.45\textwidth,trim={0 0 0 -30}, clip=false]{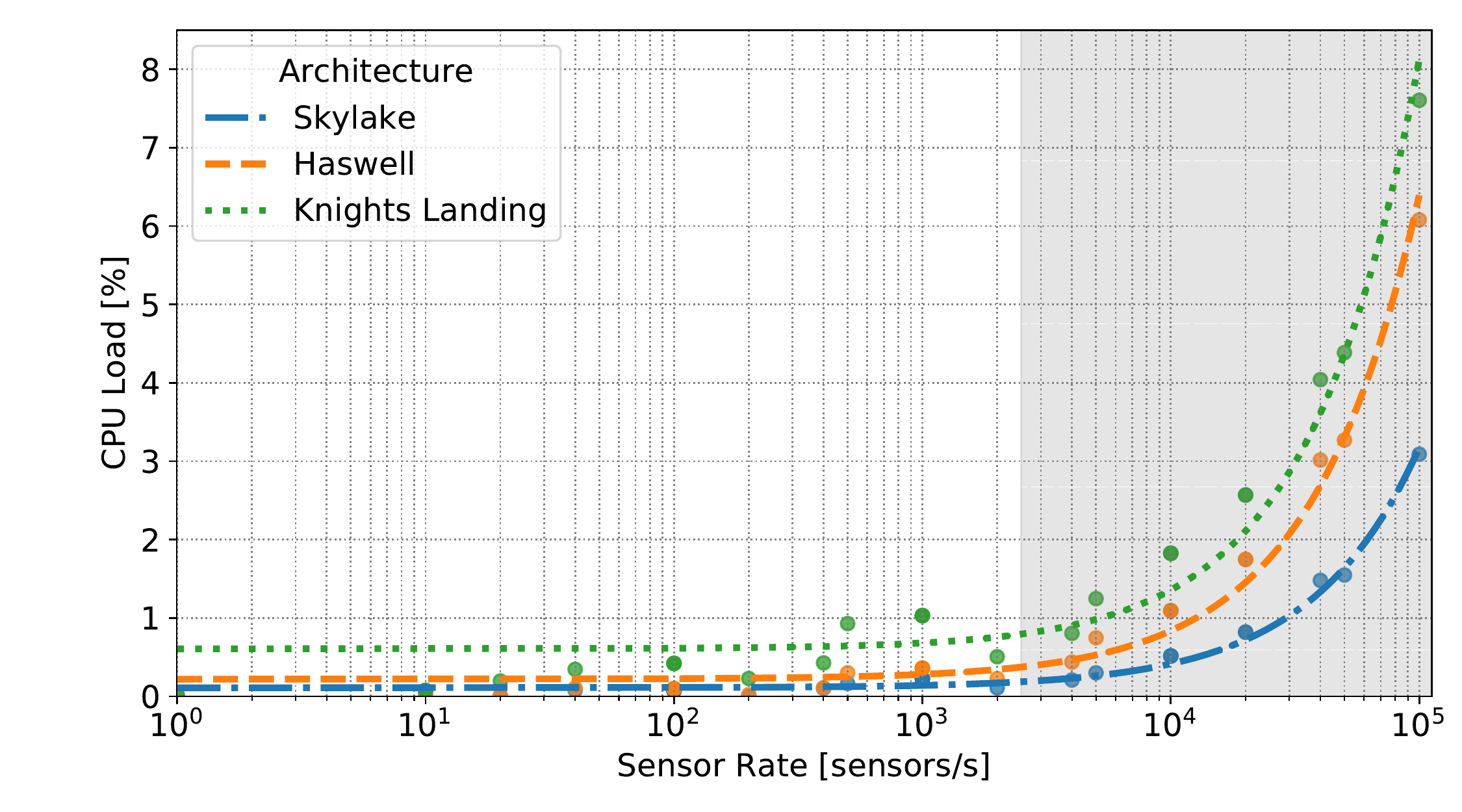}
 \caption{CPU load scaling for the three reference architectures under different sensor rates. The white area highlights typical configurations for production HPC environments.}
 \label{results:performancescaling}
\end{figure}

We now discuss a generic model to infer the performance of our Pusher solution in terms of per-core CPU load on each architecture, as a function of the sensor rate (i.e., the ratio of the number of instantiated sensors and the sampling interval). We use the performance data obtained in the experiment discussed in Section~\ref{section:testperf} on the three reference architectures. Figure~\ref{results:performancescaling} shows the observed average per-core CPU load across configurations as well as fitted curves resulting from linear regression. Since we cover a broad range of sensor rates, the X-axis is shown in logarithmic scale.

Results show that varying levels of performance are achieved across the reference architectures. The Skylake architecture, in particular, shows the best scaling curve with 3\% peak CPU load, whereas Knights Landing once again shows the worst results, with 8\% peak CPU load. In all architectures, however, CPU load is below 1\% for configurations with a sensor rate of 1,000 or less. Most importantly, Pushers follow a distinctly linear scaling curve on all architectures. This implies that system administrators can reliably infer the average CPU load of the Pusher on a certain system by means of linear interpolation, with the following equation: 

\begin{equation}
    L_p(s) = L_p(a) + (s - a)\frac{L_p(b) - L_p(a)}{b - a}
\label{eq:scalingmodel}
\end{equation}

In Equation \ref{eq:scalingmodel}, \(L_p\) represents the average CPU load, while \(s\) is the target sensor rate, and \(a\) and \(b\) are two reference sensor rates for which the average CPU load was measured. 

\subsection{Collect Agent Performance}

In this subsection we analyze the performance and scalability of our Collect Agent component, to prove its effectiveness in supporting a large-scale monitoring infrastructure. In order to evaluate its scalability we focus on the CPU load metric, as defined in Section~\ref{section:expsetup}. We do not analyze the performance of the Cassandra key-value store to which Collect Agents write, as it is a separate component. Similar to Section~\ref{section:pusherperf}, we perform tests by running Pushers with the \emph{tester} plugin under different configurations. In this test, we used a sampling interval of 1 second, and experimented with different numbers of Pushers, executed from separate nodes, each sampling a certain number of sensors.

\begin{figure}[t]
 \centering
 \includegraphics[width=0.45\textwidth,trim={0 0 0 0}, clip=true]{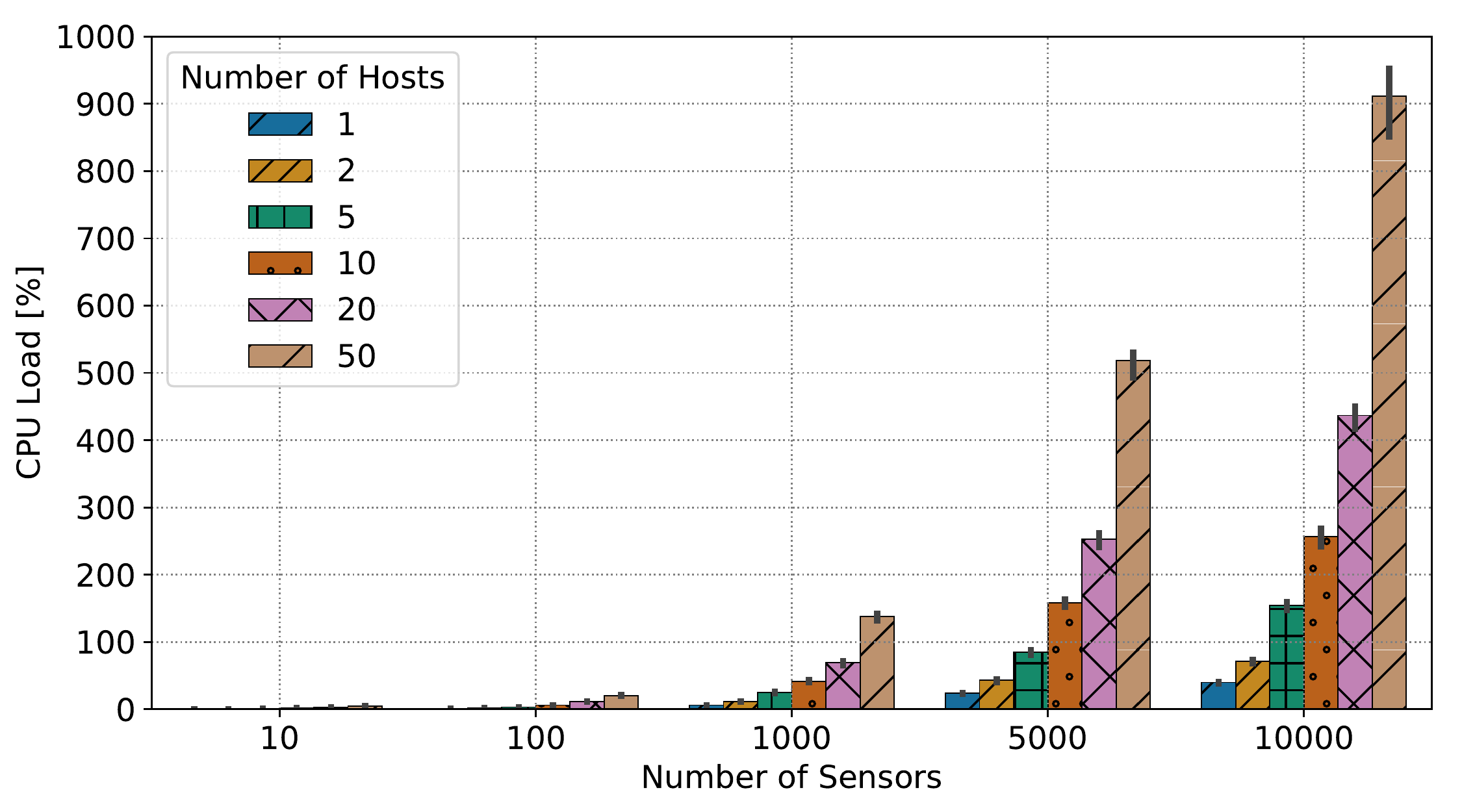}
 \caption{Average per-core CPU load of the Collect Agent under different configurations.}
 \label{results:collectResults}
\end{figure}

We show the results of our tests in Figure~\ref{results:collectResults}. In the configurations that use 1,000 sensors or less, we reach saturation of a single CPU core only with 50 concurrent hosts. In the most intensive configurations, multiple CPU cores are fully used, but even in the worst-case scenario we observe an average CPU load of 900\%, which translates into 9 fully-loaded cores: this corresponds to a Cassandra insert rate of 500,000 sensor readings per second (10,000 sensors sampled each second by 50 concurrent Pushers), which is equivalent to that of a production configuration in a medium-scale system. 

\vspace{2mm}

\section{Case Studies}
\label{section:usecase}

In the following, we illustrate two real-world case studies using \dcdb, illustrating both facility and application-level analysis on top of a single system and with shared metrics. In the first, we prove its effectiveness for monitoring and correlating infrastructure data from different sources by analyzing the cooling system's ability to remove heat, whereas in the second we focus on showing the usefulness of high-frequency monitoring data collected in compute nodes to characterize the power consumption of applications.

\subsection{Efficiency of Heat Removal}
One of the requirements in the procurement of our CooLMUC-3 system was to achieve high energy efficiency through the employment of direct warm-water cooling. Megware, the chosen system integrator, provided a 100\% liquid-cooled solution that not only liquid-cools the compute nodes, but also all other components, including power supplies and network switches. Hence, the entire system does not require any fans in the compute racks and therefore allows us to thermally insulate them. This reduces the heat emission to the compute room to close to zero. To help study its efficiency, the system is  broadly instrumented and provides a wide range of infrastructure sensors and measuring devices, such as power sensors and flow meters. We monitor these sensors and devices in \dcdb to evaluate the efficiency of the system's water cooling solution by calculating the ratio between the heat removed via warm water and the system's total electrical power consumption. 

\begin{figure}[t]
 \centering
 \includegraphics[width=0.47\textwidth,trim={0 15 0 0}, clip=false]{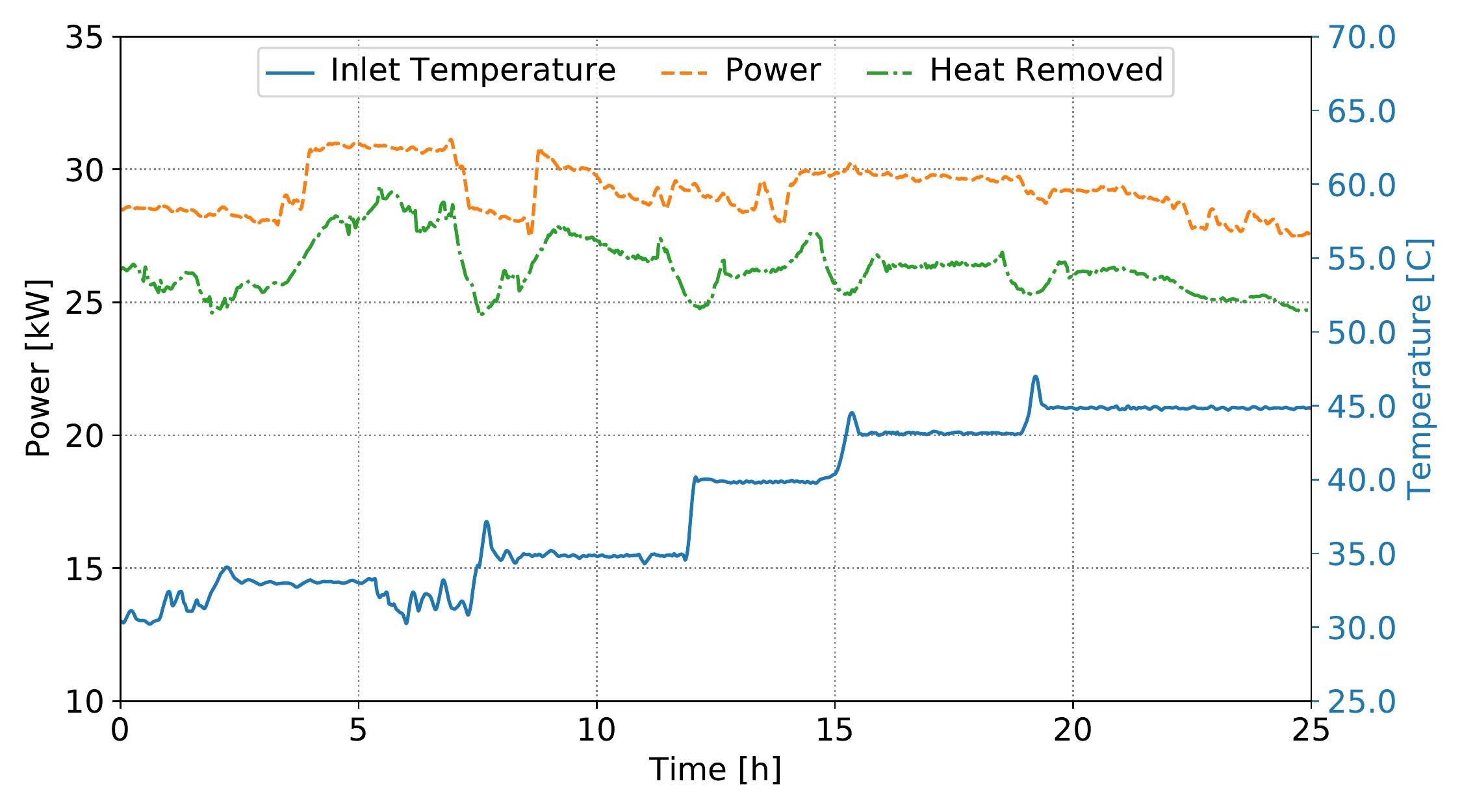}
 \caption{\dcdb Use Case 1: Efficiency of Heat Removal --- The graph shows that the heat removal efficiency of our CooLMUC-3 system is around 90\%, independent of inlet water temperature.}
 \label{results:useCase}
\end{figure}

Figure \ref{results:useCase} depicts in detail the behavior of the monitored metrics for our case study, specifically the total power consumption of the system, the total heat removed from the system by the liquid-cooling circuit, and its inlet water temperature. All data has been collected out-of-band by running one Pusher and one Collect Agent on two different management servers and by leveraging the Pusher's REST and SNMP plugins. As it may be expected, the instrumentation employs sensors only at the node or rack levels, which do not supply a picture of the entire system's status. Hence, we defined aggregated metrics in \dcdb using the virtual sensors (as described in Section~\ref{sections:architecture}), which prove to be particularly suitable for this use case. Using \dcdb, we were able to easily record all relevant sensors and to calculate the average ratio between the total heat removed and the power drawn, which turned out to be approximately 90\%, showing very high efficiency of our new system's water cooling solution. We further observe that, for rising inlet water temperatures, the gap between power and heat removed does not increase, suggesting that the insulation of the entire racks is effective in reducing the emission of heat radiation to ambient air. 

\subsection{Application Characterization}
Monitoring data can be used to implement a feedback loop inside of HPC systems, by using it to make informed and adaptive management decisions. One such use case involves using monitoring data in compute nodes to characterize the relationship between the throughput and power consumption of running applications, and thus change parameters such as the CPU frequency at runtime to improve the overall energy efficiency. In this scenario, for example, when applying monitoring data to drive \emph{Dynamic Voltage and Frequency Scaling} (DVFS), the frequency of sampling needs to be high (i.e., greater than 1Hz) so as to react quickly to the frequent changes that occur in application behavior~\cite{mittal2014power}.

\begin{figure}[t]
 \centering
 \includegraphics[width=0.47\textwidth,trim={0 0 0 0}, clip=true]{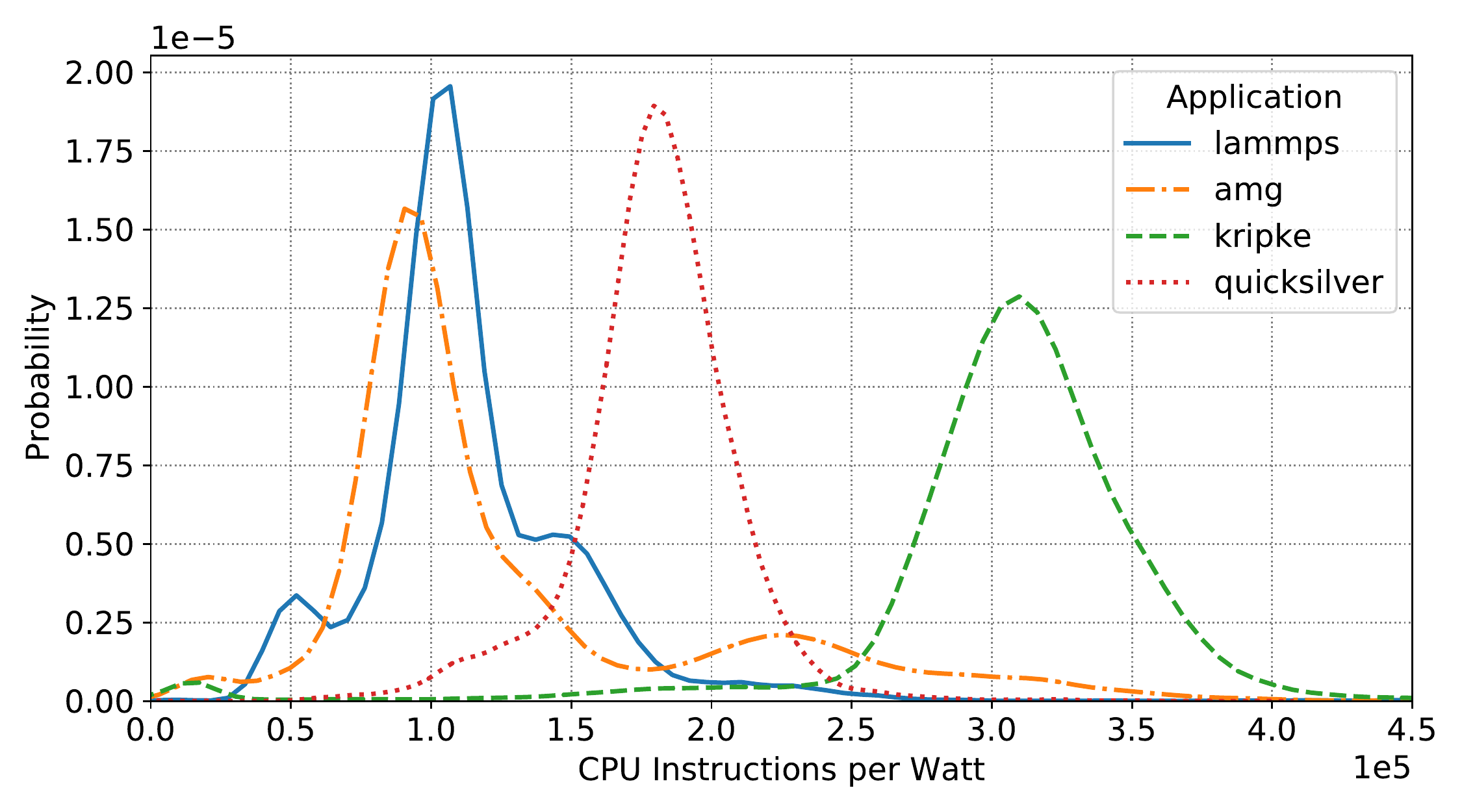}
 \caption{\dcdb Use Case 2: Application Characterization --- The plot shows the probability density functions of the per-core instructions per Watt observed for different CORAL-2 applications.}
 \label{results:useCase2}
\end{figure}

Here, we present a characterization of the four applications from the CORAL-2 suite used in Section~\ref{sections:results}. We execute several runs of the applications on a single node in our CooLMUC-3 system, while using \dcdb with a 100ms sampling interval. The application, node and \dcdb configurations are as described in Section \ref{section:expsetup}. In particular, we try to gain insight into the characteristics of each single application by analyzing the ratio between the number of per-core \emph{retired instructions} and the node's power consumption at each time point. In Figure \ref{results:useCase2} we show the fitted probability density function of the resulting time series for each application. We can see that each application shows a distinct behavior: Kripke and Quicksilver exhibit very high mean values, translating to a high computational density, while applications such as LAMMPS or AMG show lower values. Moreover, the distributions of the two latter applications show multiple trends, indicating a dynamic behavior that changes over time. Obviously, those variations vastly depend on the code profile of the underlying applications. In this context, the use of a fine-grained monitoring tool, such as \dcdb,  significantly contributes to distinguishing different application patterns and to support the implementation of optimal operational modes, leading to better system performance and efficiency (e.g., by selecting optimal CPU frequencies).

\section{Related Work}
\label{sections:relatedwork}
Many insular components for data center and HPC system management are currently available, with varying feature sets, broadness and architectures. The \emph{Examon} monitoring framework~\cite{beneventi2017continuous} shares a similar design philosophy with our monitoring solution, as it also employs a push-based monitoring model, allows for accurate fine-grained monitoring, and uses MQTT and Apache Cassandra as communication protocol and data store, respectively. Examon, however, has been mostly developed as an \emph{ad-hoc} solution for certain production environments, and lacks a modular and plugin-oriented architecture, making integration of new data sources difficult. Moreover, \dcdb provides synchronization of measurements across nodes, which helps reduce the interference on parallel applications and allows for accurate correlation of sensors. 

One of the most common open-source monitoring frameworks in the HPC domain is the \emph{Lightweight Distributed Metric Service} (LDMS)~\cite{agelastos2014lightweight}, which has been shown to be suitable for deployment on large-scale systems. However, while plugin-based, the LDMS architecture is not designed for customization. As such, development of new plugins for specific purposes requires considerable effort. Moreover, storage options for sensor data in LDMS are limited, and since the communication protocol between sampler and aggregator processes is custom, integration with other frameworks and environments is difficult. Finally, the pull-based model adopted by LDMS is problematic for fine-grained monitoring, which requires high sampling accuracy and precise timing. The \emph{Ganglia}~\cite{massie2004ganglia} and \emph{Elastic Stack}\footnote{\url{https://www.elastic.co/products/}} open-source monitoring frameworks share similar issues as LDMS, but are not designed for large-scale HPC setups.

On the data transport side, frameworks like the Multicast Reduction Network (MRNet) enable high-throughput and efficient multicast and reduction operations in distributed systems~\cite{DBLP:conf/sc/RothAM03}. In particular, MRNet relies on a tree-based overlay network for communication, whereby retrieval of data is performed from the leaves to the root of the tree. Packet aggregation can be implemented via customizable filters. While MRNet could be integrated into \dcdb for its communication, we opted to deploy an IoT-based communication solution instead, in our case MQTT, due to its wider spread use, higher acceptance among our administrators as well as a more loosely-coupled initialization, which allows us to easily deploy \dcdb beyond job boundaries. Furthermore, since the purpose of our framework is holistic continuous monitoring, filtering is not desired in our case, but was one of the major drivers behind MRNet and is one of its major advantages. 

There are many commercial and closed-source products for system-wide monitoring. Among these, the most popular are \emph{Nagios}~\cite{barth2008nagios}, \emph{Zabbix}~\cite{olups2016zabbix} and \emph{Splunk}~\cite{carasso2012splunk}, which are deployed in many data centers across the world. \emph{Icinga}\footnote{\url{https://www.icinga.com}} is a similar product, tailored for HPC systems specifically. These frameworks, however, are alert-oriented and focus on the analysis of \emph{Reliability, Availability and Serviceability} (RAS) metrics to provide insights into system behavior. They supply conventional monitoring features, but these usually focus on infrastructure-level data and cover only a very small part of the vast amount of metrics that could be monitored in a system (e.g., CPU performance counters in compute nodes).

The \emph{PerSyst} tool~\cite{Guillen2014} specializes in collecting performance monitoring data, transforming raw data into performance patterns and aggregating the data during collection. The data at the backend therefore lacks the detail of the raw performance metrics. The \emph{ScrubJay} tool~\cite{gimenez2017scrubjay} allows to automatically derive semantic relationships from raw monitored data, which is aggregated to generate performance indicators.

Performance profiling tools such as \emph{HPCToolkit}~\cite{adhianto2010hpctoolkit}, \emph{Likwid}~\cite{treibig2010likwid}, \emph{TAU}~\cite{shende2006tau} or \emph{perf}~\cite{weaver2013linux} offer extensive support for the collection and manipulation of node-level performance metrics at fine granularity, useful for application behavior characterization, but do not support the transmission and storage of monitored data nor the integration of facility data. Finally, \emph{TACC Stats}~\cite{evans2014tacc} is a comprehensive solution for monitoring and analysis of resource usage in HPC systems at multiple resolution levels, whereas tools such as \emph{Caliper}~\cite{boehme2016caliper} provide generic interfaces to enable application instrumentation. These tools are not designed to perform system-wide monitoring, but rather application analysis and have a different focus compared to our monitoring solution, yet could potentially be included in \dcdb as additional data sources.

\section{Conclusions and Future Work}
\label{sections:conclusions}

In this paper we have presented \dcdb, a novel monitoring framework for HPC systems that is designed to be modular, scalable and easily customizable. Our framework supports the most common standards and protocols for the collection of data in HPC systems, covering a broad range of performance events and sensor types that can be monitored. We characterized the footprint and overhead of \dcdb, which was observed to be very small, by evaluating it on several production HPC systems, with different architectures and scales. We observed overhead against state-of-the-art benchmarks to also be very low and on par with other monitoring solutions. We already deployed \dcdb in the context of several HPC research projects with successful results, and we expect to complete soon its deployment on all production HPC systems at LRZ.

As future work, we plan to further extend \dcdb and develop further plugins in order to support a broader range of sensors and performance events, such as those deriving from GPU usage. Also, since application data in \dcdb is currently only extracted from data sources such as CPU performance counters, we plan to implement plugins to collect profiling data as well, so as to extend the application analysis capabilities of \dcdb. Moreover, we plan to implement a \emph{streaming data analytics} layer highly-integrated in our framework, which will offer novel abstractions to aid in the implementation of algorithms for many data analytics applications in HPC, such as energy efficiency optimization or anomaly detection. This framework will be able to fetch live sensor data and perform online data analytics at the Collect Agent or Pusher level, and it will make use of features in our monitoring solution such as sensor caching, RESTful APIs, and the Pusher's plugin-based architecture.

\begin{acks}
The research leading to these results has received funding from the Mont-Blanc 1 and 2 projects and the DEEP project respectively under EU FP7 Programme grant agreements n. 288777, 61042 and 287530, and from the DEEP-EST project under the EU H2020-FETHPC-01-2016 Programme grant agreement n. 754304.
\end{acks}

\bibliographystyle{ACM-Reference-Format}
\bibliography{main}
\end{document}